\documentclass[twoside,12pt]{article}
\usepackage{epsfig}

\def\Journal#1#2#3#4{{#1} {#2} (#4) #3 }

\def\NPA{{\em Nucl. Phys.} A}

\def\PRL{\em Phys. Rev. Lett.}

\def\PRC{{\em Phys. Rev.} C}

\newcommand{\be}{\begin{equation}}
\newcommand{\ee}{\end{equation}}
\newcommand{\bea}{\begin{eqnarray}}
\newcommand{\eea}{\end{eqnarray}}
\newcommand{\nn}{\nonumber}

\newcommand{\Image}{\mathop{\rm Im}\nolimits}
\newcommand{\sump}{\mathop{{\sum}'}}
\newcommand{ \bb }{$2\nu\beta\beta$}
\newcommand{ \bbm }{$2\nu\beta^-\beta^-$}
\newcommand{ \bbb }{$0\nu\beta\beta$}
\begin{document}

\title{ \vspace{1cm} First application of the continuum-QRPA 
to description of the double beta decay}
\author{Vadim Rodin$^{\,1}$ and Amand Faessler$^{\,1}$
\\
$^1$Institute for Theoretical Physics, University of Tuebingen, Germany}

\maketitle

\begin{abstract} 
A continuum-QRPA approach to calculation of the \bb- and \bbb-amplitudes has been formulated.
For $^{130}$Te a regular suppression (about 20\%) of the high-multipole contributions to the \bbb-amplitude
has been found which can be associated with additional ground state correlations appearing from the transitions to 
collective states in the continuum. At the same time the total calculated 
\bbb-amplitude for $^{130}$Te gets suppressed by about 20\% as compared to the result of 
the usual, discretized, QRPA.
\end{abstract}

\section{Introduction}

At present, the most elaborated analysis 
of the uncertainties in the \bbb -decay nuclear matrix elements $M^{0\nu}$ calculated within the QRPA and RQRPA
has been performed in the recent works~\cite{Rod03a,Rod05}. The experimental \bbm-decay rate 
has been used to adjust the most relevant parameter, the strength $g_{pp}$ of the particle-particle
interaction. With such a procedure the values of $M^{0\nu}$ have been shown to become essentially independent 
on the size of the single-particle (s.p.) basis. 
Furthermore, the $M^{0\nu}$'s have been demonstrated to be also rather stable with respect  
to the possible quenching of the axial vector strength %parameterized by reducing the coupling constant 
$g_A$,  as well as to the uncertainties of parameters describing the short range nucleon correlations.  

The calculations in~\cite{Rod03a,Rod05} have been performed within ``the standard QRPA" scheme in which the BCS
ground state and the spectrum of the excited states are built on a discrete s.p. basis.
However, in order to really assess the uncertainty of the QRPA calculations of the $M^{0\nu}$ one has to address the questions regarding the accuracy of ``the standard QRPA" itself, and whether there are not some important contributions missing.

The accepted procedure~\cite{Rod03a,Rod05} of averaging the results of the calculations 
performed with different choices of the model space size looks rather %weird.
weakly justified.
One can expect {\it a priori} that enlargement of the model space should lead to more accurate 
results (in other words, any basis truncation leads to an uncertainty).
The usual statement that ``in the QRPA one can include essentially an unlimited set of
single-particle states"~\cite{Rod03a} can be considered at presence only as principally correct, but not realized 
in practice.
In reality, the large number of the major shells $N\gg 1$ in the QRPA
calculations can  only be achieved by adding low-lying major shells composed of the bound states.
Basically, only one major shell lying higher than the Fermi-shell one can be considered, because 
one immediately encounters principal limitations of the approximation 
of the single-particle continuum by discrete levels.
Neglecting the single-particle continuum leads to a missing strength in the usual QRPA calculations, especially for description of the high-multipole excitations with 
$L\ge 2$~\footnote{Note, that the QRPA is barely suitable for the description of 
the multipole contributions to $M^{0\nu}$ with $L\ge 5$ because 
they are completely dominated by the short-range behavior of the wave function.}. 

The contribution of these multipoles to $M^{0\nu}$ is particularly important
because the monopole (Fermi + Gamow-Teller) one is suppressed due to the symmetry
constrains (see, e.g., the multipole decomposition of $M^{0\nu}$ in Fig.~5 of~\cite{Rod05};
for a recent discussion how the SU(4)-symmetry violation by the particle-particle
interaction affects $M^{2\nu}$ see~\cite{Rodin05}).
Furthermore, the pure pairing contribution to $M^{0\nu}$ is almost completely (by an order of magnitude) 
suppressed by the ground state correlations, short-range correlations etc.,
therefore fine effects can be expected to play an important role.

The question about the dependence of the QRPA results on the s.p.-basis size as a source of the uncertainties 
in the calculated $M^{0\nu}$'s would be completely resolved if one could include 
the entire s.p. basis into the calculation scheme. The only possible way principally to perform
such {\em ultimate-basis QRPA calculations} is provided within the continuum-QRPA. 
Also, to have an alternative formulation of the QRPA can help to promote our understanding 
of the current QRPA results and their deficiencies to a higher level. 
In particular, the continuum-QRPA provides a regular way of using realistic wave functions of the continuum states
in terms of the Green's functions and there is no need to approximate them by the oscillator ones.

Two principal effects of the inclusion of the s.p. continuum within the pn-QRPA, which affect $M^{0\nu}$'s 
in an opposite way, can be expected. 
First, pairing in the continuum can increase respective $0\nu\beta\beta$ sum rules (%$M^{0\nu}\,\uparrow$, 
the  smallness of the $\Delta/E$ in the continuum can be compensated by a large corresponding
partial s.p. matrix element). Second, additional g.s. correlations can appear due to collective multipole states 
in the continuum that decreases $M^{0\nu}$.
It is noteworthy that at the moment the continuum-QRPA consistently including pairing in
the continuum has not been formulated yet and only the continuum-QRPA with the pairing
realized on a discrete basis can be used.

The principal aim of this work is to formulate for the first time a method to calculate the double beta decay matrix elements within the continuum-QRPA.

\section{Continuum-QRPA}

The continuum-RPA has been used for a long time
to successfully describe structure and decay properties of various giant resonances
and their high-lying overtones embedded in the single-particle continuum.
However, to apply the continuum-RPA in open-shell nuclei one has to take the nucleon pairing into consideration
and develop a continuum-QRPA approach. The approach should account for the important effects of
particle-particle interaction along with the usual particle-hole one. Such a continuum-QRPA approach
has been developed in~\cite{Bor90,Rod03}.
In~\cite{Bor90} the approach has been applied to the analysis of
the low-energy part of the Gamow-Teller (GT) strength distribution for description of the single-beta decay relevant
to astrophysical applications.

\subsection{\it QRPA equations in the coordinate representation}

The system of homogeneous equations for the forward and backward amplitudes
$X^{(J^\pi s)}_{\pi\nu}$ and $Y^{(J^\pi s)}_{\pi\nu}$,
%$X^{J}_{\pi\nu}(s)=\langle s,J\mu | \left ( A^{J\mu}_{\pi\nu} \right )^+ |0\rangle$
%and $Y^{J}_{\pi\nu}(s)=\langle s,J\mu |\tilde A^{J\mu}_{\pi\nu} |0\rangle$
%
%\begin{eqnarray}
%&|JM  \rangle = { Q^{\dagger}_{JM }}|0^+_{RPA}\rangle  \qquad  \qquad
%{  Q^{\dagger}_{JM}} =\displaystyle \sum\limits_{\pi\nu}
% \left [ X_{\pi\nu} {  A^\dagger_{\pi\nu}} - Y_{\pi\nu}{  \tilde{A}_{\pi\nu}}\right ] \nonumber\\
%\nonumber \end{eqnarray}
%
respectively,
is usually solved to calculate the energies $\omega_s$ and the wave functions
$|J\mu , s\rangle$ of the isobaric nucleus within %the quasiboson version of 
the pn-QRPA (see, e.g.,~\cite{fae98}).
It is impossible to handle the infinite number of the amplitudes $X,Y$ if
one wants to include the single-particle continuum. Instead, by going into the coordinate representation 
the pn-QRPA can be reformulated in equivalent terms of four-component radial transition density 
$\{\varrho^{(J^\pi s)}_{I}\}\ (I=1,\dots 4)$. The components are determined by 
the standard QRPA amplitudes $X$ and $Y$ as follows:

\bea
&& \varrho^{(J^\pi s)}_{I}(r)=\sum\limits_{\pi\nu} R^{(J^\pi s)\pi\nu}_I \, 
\chi_{\pi\nu}(r),\label{defvarrho}\\
&&\left({\begin{array}{c}
R^{\pi\nu}_{p-h}\\
R^{\pi\nu}_{h-p}\\
R^{\pi\nu}_{p-p}\\
R^{\pi\nu}_{h-h}
\end{array}}\right)
=
\left({\begin{array}{c}
u_\pi v_\nu X_{\pi\nu}+v_\pi u_\nu Y_{\pi\nu}\\
u_\pi v_\nu Y_{\pi\nu}+v_\pi u_\nu X_{\pi\nu}\\
u_\pi u_\nu X_{\pi\nu}-v_\pi v_\nu Y_{\pi\nu}\\
u_\pi u_\nu Y_{\pi\nu}-v_\pi v_\nu X_{\pi\nu}
\end{array}}\right)
\nn
\eea
where $u,~v$ are the coefficients of Bogolyubov transformation,
$\chi_{\pi\nu}(r)=t^{(J)}_{(\pi)(\nu)}\,\chi_\pi(r)\chi_\nu(r)$ 
with $t^{(J)}_{(\pi)(\nu)}=\frac{1}{\sqrt{2J+1} }\langle\pi\|T_{JLS}\|\nu\rangle$
being the reduced matrix element of the spin-angular tensor $T_{JLS\mu}$ and $\chi_{\pi}(r)$ ($\chi_{\nu}(r)$)
being the radial wave function of a single-particle proton (neutron) state. Hereafter 
we shall systematically omit the superscript ``${J^\pi s}$" when it does not lead to a confusion.

According to the definition (\ref{defvarrho}), the elements
$\varrho_1,\varrho_2,\varrho_3,\varrho_4$ can be called the particle-hole, hole-particle, hole-hole and
particle-particle components of the transition density, respectively.
%, which can be generally considered as a 4-dimensional vector:$\{ \varrho^{J}_i\}$. 
In particular, the transition matrix element to the state $|s,J\mu\rangle$ corresponding
to a probing particle-hole operator $\hat V^{(-)}_{J\mu}=\sum_a V_{J}(r_a)T_{JLS\mu}\ \tau^-_a$
is determined by the element $\varrho_1$:\ \ $\int \varrho^{(J^\pi,s)}_1(r) V_{J}(r)\, dr$.

The pn-QRPA system of equations for the elements $\varrho_i^{(J^\pi s)}$ is as follows:

\be
\varrho^{(J^\pi s)}_{I}(r)=
\sum\limits_{K} \int A^{(J^\pi)}_{IK}(rr',\omega=\omega_s)\, F^{(J^\pi)}_K(r'r'')\, \varrho^{(J^\pi s)}_{K}(r'')
\, dr'dr'',\label{eq1varrho}
\ee
or schematically, denoting all the integrations and summations as $\{\dots\}$, 
$\varrho=\{AF\varrho\}$,
where $F^{(J^\pi)}_K(r_1r_2)$ represents the residual interaction in $K$-channel 
($K=1,2$ --- p-h channel, $K=3,4$ --- p-p channel) after separation of the spin-angular variables.
%According to Eq.~(\ref{eqvarrho}), 
The $4\times 4$ matrix $A_{IK}(r_1r_2,\omega)$ is the radial part
of the free two-quasiparticle propagator.
%, whereas the quantities $v_{K}(r_1)=\int F_K(r_1r_2)\varrho_{K}(r_2)\,dr_2$ are the
%elements of the radial transition potential.
The expressions for the elements of the free two-quasiparticle propagator $A_{IK}$
can be obtained by making use of the regular and anomalous single-particle Green's
functions for Fermi-systems with nucleon pairing in an analogous way to how it was done
in the monograph~\cite{Mig83} to describe the Fermi-system response to a
single-particle probing operator acting in the neutral channel~\cite{Rod03}.
%Namely, the matrix $A$ can be depicted as the set of the following diagrams:
%
%\vskip-0.5cm
%\mbox{
%\centerline{\includegraphics[width=.5\textwidth,height=.3\textheight]{propag.eps}}
%}
%\noindent Here upper (lower) lines correspond to the propagation of the proton
%(neutron) quasiparticles. 
The corresponding analytical representations for the elements $A_{IK}(r_1r_2,\omega)$are:

\bea
&A_{IK}(r_1r_2,\omega)=\sum\limits_{\pi\nu} 
\, \chi_{\pi\nu}(r_1)\chi_{\pi\nu}(r_2)\, A^{\pi\nu}_{IK}(\omega); \ \ \ \ A^{\pi\nu}_{IK}=A^{\pi\nu}_{KI}\label{A} \\
& A^{\pi\nu}_{11}=\displaystyle\frac{u^2_\pi v^2_\nu}{\omega-E_\pi-E_\nu} + \frac{u^2_\nu v^2_\pi}{-\omega-E_\pi-E_\nu},
\ \ A^{\pi\nu}_{33}=\frac{u^2_\pi u^2_\nu}{\omega-E_\pi-E_\nu} + \frac{v^2_\nu v^2_\pi}{-\omega-E_\pi-E_\nu},
\nonumber\\
%& \nonumber\\
&A^{\pi\nu}_{12}=A^{\pi\nu}_{34}=\displaystyle\frac{u_\pi v_\pi v_\nu u_\nu}{\omega-E_\pi-E_\nu} + 
\frac{u_\pi v_\pi v_\nu u_\nu}{-\omega-E_\pi-E_\nu}\nonumber\\
&A^{\pi\nu}_{13}=u_\nu v_\nu \displaystyle(\frac{u^2_\pi}{\omega-E_\pi-E_\nu} - \frac{v^2_\pi}{-\omega-E_\pi-E_\nu}),\ \
A^{\pi\nu}_{14}=u_\pi v_\pi (\frac{v^2_\nu}{\omega-E_\pi-E_\nu} - \frac{u^2_\nu}{-\omega-E_\pi-E_\nu})\ ,\nonumber\\
& A^{\pi\nu}_{22}(\omega)=A^{\pi\nu}_{11}(-\omega),\ \ A^{\pi\nu}_{44}(\omega)=A^{\pi\nu}_{33}(-\omega), \ \ A^{\pi\nu}_{23}(\omega)=A^{\pi\nu}_{14}(-\omega),\ \ A^{\pi\nu}_{24}(\omega)=A^{\pi\nu}_{13}(-\omega).
\nonumber
\eea

The total two-quasiparticle propagator $\tilde A$ that includes the QRPA iterations of the p-h and p-p interactions satisfies a Bethe-Goldstone-type integral equation:
\be
	\tilde A=A+\{AF	\tilde A\} \nonumber
\ee
and has the following spectral decomposition:
\bea
&\displaystyle	\tilde A^{(J^\pi)}_{IK}(r_1r_2,\omega)=
\sum_s \frac{\varrho^{(J^\pi s)}_{I}(r_1)\varrho^{(J^\pi s)}_{K}(r_2)}{\omega-\omega_s+i\delta}
-\sum_s \frac{\varrho^{(J^\pi s)}_{I}(r_1)\varrho^{(J^\pi s)}_{K}(r_2)}{\omega+\omega_s-i\delta}
\label{spectrA}
\eea

Thus, all the necessary information about the QRPA solutions resides
%is contained 
in the poles of $\tilde A^{(J^\pi)}$.

\subsection{\it Strength functions\label{SSF}}

One can use $\Image\tilde A$ to calculate different strength functions. 
Strength function corresponding to a charge-exchange single-particle probing operator
\be
\hat V^{(\mp)}_{J\mu}=\sum_a V_{J}(r_a)T_{JLS\mu}(n_a) \tau_a^{(\mp)}
\ee
acting in $\beta^{(\mp)}$-cnannel is defined by the usual expression:
\be
S^{(\mp)}(\omega)=\sum\limits_{s} \left | \langle s | \hat V^{(\mp)}_{J\mu} |0\rangle\right |^2
\delta(\omega-\omega_s^\mp)\nonumber
\ee
with $\omega_s^\mp=E_s^\mp-E_0$ being the excitation energy of the corresponding
isobaric nucleus measured from the ground state of the parent nucleus.
Making use of the spectral decomposition (\ref{spectrA}) one can easily verify the following
integral representations of the strength functions in term of  $\Image \tilde A$:
\bea
&S^{(-)}(\omega^-)=-\frac 1\pi \Image \int V_{J}(r_1) \tilde A_{11}^{(J^\pi)}(r_1r_2;\omega) V_{J}(r_2)\, dr_1dr_2 \\
&S^{(+)}(\omega^+)=-\frac 1\pi \Image \int V_{J}(r_1) \tilde A_{22}^{(J^\pi)}(r_1r_2;\omega) V_{J}(r_2)\, dr_1dr_2
\eea
or, schematically, 
\bea
&S^{(-)}_V(\omega^{-})=-\frac 1\pi \left\{V \tilde A_{11}(\omega) V \right\}\ \ \ \ 
 S^{(+)}_V(\omega^{+})=-\frac 1\pi \left\{V \tilde A_{22}(\omega) V \right\}
\nonumber
\eea
with $\omega^\mp=\omega\pm (\lambda_p-\lambda_n)$. The calculated pn-QRPA spectrum in $\omega$ is
to be shifted in energy in order to be measured from the ground state of the parent nucleus.
It has to do with the fact that in the QRPA the BCS hamiltonian $\hat H - \lambda_p\hat Z - \lambda_n\hat N$ is used.

One can also define a non-diagonal strength function like 
\be
S^{(- -)}_V(\omega)=\sum\limits_{s}  \langle 0' | \hat V^{(-)}_{J\bar\mu} |s \rangle 
\langle s | \hat V^{(-)}_{J\mu} |0\rangle \delta(\omega -\omega'_s)\label{Smm}
\ee
with $\omega'_s=E_s-(E_0+E_{0'})/2$. Such a strength function is closely 
related to the amplitude of the \bb-decay. To calculate $S^{(- -)}_V(\omega)$
within the pn-QRPA one faces the usual problem that the spectrum $|s\rangle$ comes out
slightly different when calculated starting from the initial or final ground states.
Identifying the BCS vacuum $|0'\rangle$ with that of $|0\rangle$ one gets
\be
S^{(- -)}(\omega^-)=-\frac 1\pi \Image\int V_{J}(r_1) \tilde A_{12}^{(J^\pi)}(r_1r_2;\omega) V_{J}(r_2) \, dr_1dr_2
\ee
or, alternatively, identifying $|0\rangle$ with $|0'\rangle$
\be
S^{(- -)}(\omega^+)=-\frac 1\pi \Image\int V_{J}(r_1)\tilde {A'}^{(J^\pi)}_{12}(r_1r_2;\omega) V_{J}(r_2) \, dr_1dr_2
\ee
where $	\tilde A'$ is calculated with respect to $|0'\rangle$.

To calculate the strength functions, it is more convenient 
to use a system of inhomogeneous cQRPA equations in $\beta^-$-channel: 
\bea
&S^{(-)}_V(\omega^{-})=-\frac1\pi\Image \sum\limits_{K} \int V(r_1)\,
A_{1K}(r_1r_2,\omega)\,\tilde V_{K}(r_2,\omega)\, dr_1dr_2, \label{SFm}\\
&S^{(--)}_V(\omega^{-})=-\frac1\pi\Image \sum\limits_{K} \int V(r_1)\,
A_{2K}(r_1r_2,\omega)\,\tilde V_{K}(r_2,\omega)\, dr_1dr_2, \label{SFmm}\\
&\tilde V_{I}(r,\omega)=V(r)\delta_{I1}+
\sum\limits_{K} \int F_K(rr_1)\, A_{IK}(r_1r_2,\omega)\,\tilde V_{K}(r_2,\omega)\, dr_1dr_2,
\label{Veff}
\eea
or, in $\beta^+$-channel:
\bea
&S^{(+)}_V(\omega^{+})=-\frac1\pi\Image \sum\limits_{K} \int V(r)\, A_{2K}(r_1r_2,\omega)\,\tilde V_{K}(r_2,\omega)
\, dr_1dr_2, \label{SFp}\\
&\tilde V_{I}(r,\omega)=V(r)\delta_{I2}+
\sum\limits_{K} \int F_K(rr_1)\, A_{IK}(r_1r_2,\omega)\,\tilde V_{K}(r_2,\omega)\, dr_1dr_2.
\label{Veff1}
\eea

\subsection{\it Taking the s.p. continuum into consideration}

Up to now the way of taking the s.p. continuum into consideration
has not been specified explicitly within this formulation of the pn-QRPA.
If one lets the double sums in (\ref{A}) run just over the bound proton and neutron s.p. states, 
the version of the pn-QRPA presented in two preceeding sections is fully equivalent to the usual discretized one
formulated in terms of $X$ and $Y$ amplitudes. We make use of this fact and compare discrete-QRPA results 
calculated in these two different, but formally equivalent, ways in order to check the consistency of the scheme.

To take the s.p. continuum into consideration, one has to do the following 
%to the quantities entering the sums 
in (\ref{A}):
\begin{enumerate}

\item To approximate $v_i,\ u_i$ and $E_i$ by their no-pairing values $v=0(1),\ u=1(0)$, $E=|\varepsilon-\lambda|$ for those s.p. states which lie far from the chemical potential
(i.e. $|\varepsilon-\lambda|\gg \Delta$). The accuracy of this approximation is 
$%\left ( 
\frac{\Delta}{|\varepsilon-\lambda|}
%\right)^2
$.

\item To use the s.p. Green function: $g_{(\alpha)}(r_1r_2,\varepsilon) =\displaystyle\sum_\alpha\frac{\chi_\alpha(r_1)
\chi_\alpha(r_2)}
{\varepsilon-\varepsilon_\alpha}$ to explicitly perform the sum over the s.p. states in the continuum.
\end{enumerate}
 
As an example of such an approach we present here the final expression for $A_{11}$:

\bea
A_{11}(r_1r_2,\omega)&=&
%\sum\limits_{\pi\nu}\chi_{\pi\nu}(r_1)\chi_{\pi\nu}(r_2)\,
%\left (\frac{u^2_\pi v^2_\nu}{\omega-E_\pi-E_\nu} - \frac{u^2_\nu v^2_\pi}{\omega+E_\pi+E_\nu}\right )\nonumber\\
%&& \nonumber\\
%&=&
\sum\limits_{\nu} \sump\limits_{\pi}\frac{v^2_\nu u^2_\pi\chi_{\pi\nu}(r_1)\chi_{\pi\nu}(r_2)}{\omega-E_\pi-E_\nu}
+\sump\limits_{\nu} \sum\limits_{\pi<\pi_{min}}\frac{v^2_\nu \chi_{\pi\nu}(r_1)\chi_{\pi\nu}(r_2)}{\omega-E_\nu+\lambda_\pi-\varepsilon_\pi} \nonumber\\
&&+\sum\limits_{\nu(\pi)}\left(t^{(J)}_{(\pi)(\nu)}\right )^2\chi_\nu(r_1)\chi_\nu(r_2)~v^2_\nu
g'_{(\pi)}(r_1r_2,\lambda_\pi+\omega-E_\nu)\nonumber\\
%&&+\sump\limits_{\nu} \sum\limits_{\pi<\pi_{min}}\chi_{\pi\nu}(r_1)\chi_{\pi\nu}(r_2)\frac{v^2_\nu}
%{\omega-E_\nu+\lambda_\pi-\varepsilon_\pi}\nonumber\\
&&+ \left \{\pi\leftrightarrow\nu, \omega\to -\omega \right \}
\eea
where the projected s.p. Green's function is 
$g'_{(\pi)}(r_1r_2,\varepsilon)=g_{(\pi)}(r_1r_2,\varepsilon)-\displaystyle\sump\limits_{\pi}\frac{\chi_\pi(r_1)\chi_\pi(r_2)}
{\varepsilon-\varepsilon_\pi}$. The primed sum $\sump$ runs over only those s.p. states which comprise
the BCS basis (for instance, all proton s.p. states $\pi_{min}\le \pi \le \pi_{max}$).

This continuum-QRPA method has been applied in our recent paper~\cite{Rod03} to
describe the Fermi and Gamow-Teller strength distributions in semi-magic nuclei.
%within a wide excitation-energy interval.

\subsection{\it Description of the $\beta\beta$-decay within the cQRPA}

The spectral decomposition of $\tilde A$ (\ref{spectrA}) can be used for calculation of
$\beta\beta$-decay matrix elements in a way similar to the one described in
Sec.~\ref{SSF}. For instance, the \bb-amplitude can be calculated according to the following 
expression: 
\bea
&&M^{2\nu}_{GT}=-\frac 32 \int\tilde A^{(1)}_{12}(r_1r_2;\omega=0) \, dr_1dr_2\ \ +\ \ \delta M^{2\nu}_{GT}\label{M2nu1}\\
&&\delta M^{2\nu}_{GT}=\frac {3\delta E}{\pi}\int d\omega \frac{\int \Image\tilde A_{12}^{(1)}(r_1r_2;\omega)\, dr_1dr_2}{(\omega+\delta E)\omega} 
\nonumber
\eea
where $\delta E=Q_{\beta\beta}/2+m_ec^2+\lambda_p-\lambda_n$. We use in deriving (\ref{M2nu1}) the approximation that the BCS vacuum $|0'\rangle$ of the final g.s. is taken to be the same as $|0\rangle$ of the initial g.s.
The expression~(\ref{M2nu1}) can be further rewritten in terms of the effective field $\tilde V_{K}$ (\ref{Veff}):
\bea
&&M^{2\nu}_{GT}=-\frac 32 \sum_K\int A^{(1)}_{2K}(r_1r_2;\omega=0) \tilde V_{K}(r_2;\omega=0)\, dr_1dr_2\ \ +\ \ \delta M^{2\nu}_{GT}\label{M2nu2}\\
&&\delta M^{2\nu}_{GT}=\frac {3\delta E}{\pi}\int d\omega \frac{\sum_K\int \Image A_{2K}^{(1)}(r_1r_2;\omega)\tilde V_{K}(r_2;\omega)\, dr_1dr_2}{(\omega+\delta E)\omega} \nonumber
\eea
 
The same procedure can be applied to calculate within the cQRPA the matrix element of a two-body operator
\be
\displaystyle\hat V^{(- -)}_2=\sum_{ab}\sum_{JLS\mu} V_{JL}(r_a,r_b)T_{JLS\mu}(n_a)T^*_{JLS\mu}(n_b)\tau_a^{(-)}\tau_b^{(-)}
\label{V2}
\ee
between the ground states $|0\rangle$ and $|0'\rangle$ as a sum of all partial contributions $M^{(JL)}$:
\bea
&M^{(- -)}=\langle 0' | \hat V^{(- -)}_2 |0\rangle =\displaystyle\sum_{JL} M^{(JL)}\\
&M^{(JL)} = -\frac{(2J+1)}{\pi} \int d\omega\ \int V_{JL}(r_1,r_2) \Image\tilde A^{(J^\pi)}_{12}(r_1r_2;\omega) \, dr_1dr_2
\nonumber
\eea
(the identification of the ground states described above has to be done).

The neutrino potential $\hat V^{(- -)}_2$ in the simpliest (but rather rough) Coulomb approximation
%$O_K(|\vec r_1-\vec r_2|;E_J)=|\vec r_1-\vec r_2|^{-1}$
has the well-known partial radial components
$V_{JL}(r_1,r_2)=\frac{4\pi}{2L+1}\frac{1}{r_>}\left(\frac{r_<}{r_>}\right)^L$ 
(%$J=L$, 
$r_<=min(r_1,r_2),\ r_>=max(r_1,r_2)$).
When the Jastrow factor (to account for the short range correlations)
and the energy dependence of the neutrino propagator
are considered, the decomposition of the neutrino potential over the Legandre polinomials (\ref{V2}) can be 
done numerically.

\section{First calculation results}

For the first calculations of $M^{2\nu}$ and $M^{0\nu}$ within the continuum-QRPA we adopt
a rather simple nuclear Hamiltonian similar to that used in~\cite{vog86,Engel88}.
%The Hamiltonian contains 
The chosen nuclear mean field $U(x)$ consists of the
phenomenological isoscalar part $U_0(x)$ along with the isovector $U_1(x)$
and the Coulomb $U_C(x)$ parts, both calculated consistently in
the Hartree approximation (see~\cite{Rod03}). 
The residual particle-hole interaction 
%(Landau-Migdal forces~\cite{Mig83}) 
as well as the particle-particle interaction in both 
the neutral (pairing) and charge-exchange channels are chosen in the form of the 
zero-range, $\delta$-functional, forces 
(hereafter all the strength parameters
of the residual interactions are given in units of 300 MeV$\cdot$\,fm$^3$).

\begin{table}[h]
\begin{center}
\begin{minipage}[t]{16.5 cm}
\caption{The fitted parameters $f^1_{ph}$, $g^1_{pp}$ and 
the calculated $M^{0\nu}$ for $^{130}$Te ($g_A=1.25$).}
\label{tab1}
\end{minipage}
\begin{tabular}{|c|c|c|c|}
\hline
QRPA & $f^1_{ph}$ & $g^1_{pp}$& $M^{0\nu}$ \\
\hline
discrete & 0.60& 1.20 & 2.24\\
continuum & 0.65& 1.15 & 1.79\\
\hline
\end{tabular}
\begin{minipage}[t]{16.5 cm}
\vskip 1.cm
\noindent
\end{minipage}
\end{center}
\end{table}

Fixing the model parameters is done as follows: 
\begin{itemize}
\item The pairing strengths $g^{pair}_n$, $g^{pair}_p$ are fixed within the BCS model 
to reproduce the experimental pairing energies.
\item The p-h isovector strength $f^0_{ph}$ is chosen equal to unity, $f^0_{ph}=1.0$
that allows to reproduce the experimental nucleon binding energies for closed-shell nuclei
provided the isospin-selfconsistency of the isovector p-h interaction and the symmetry potential 
$U_1(x)$ of the mean field is used (see~\cite{Rod03}).
\item The p-h spin-isovector strength $f^1_{ph}$ is fitted to reproduce the experimental
energy of the GTR.
\item  By choosing the p-p isovector strength $g^0_{pp}=(g^{pair}_n+g^{pair}_p)/2$ we restore
approximately the isospin-selfconsistency of the total residual p-p interaction.
\item The p-p spin-isovector strength $g^1_{pp}$ is chosen to reproduce the experimental value of $M^{2\nu}$.
\end{itemize}

We perform the first calculations of $M^{2\nu}$ and $M^{0\nu}$ within the continuum-QRPA for $^{130}$Te.
Also we compare the obtained results with those calculated within the usual, discretized, version of the QRPA in order to 
to see the influence of the single-particle continuum.
The chosen BCS basis contains 22 levels (oscillator shells $N=1\div 5$) that includes all bound s.p. states for neutrons
and all bound s.p. states along with 6 quasistationary states for protons.
The fitted values of the strength parameters $f^1_{ph}$ and $g^1_{pp}$ are given in Table I for both 
discretized and continuum version of the QRPA.

In Fig.~1 the calculated $g^1_{pp}$-dependence of $M^{2\nu}$ is plotted.
Note, that both $\beta^-$ and $\beta^+$ branches to construct the \bb-amplitude are calculated for $^{130}$Te,
%in the same, initial for the \bb-decay, nucleus, 
so we adopt here the same approximation as in~\cite{vog86,Engel88}.

\begin{figure}[tb]
\begin{center}
\begin{minipage}[t]{8 cm}
\epsfig{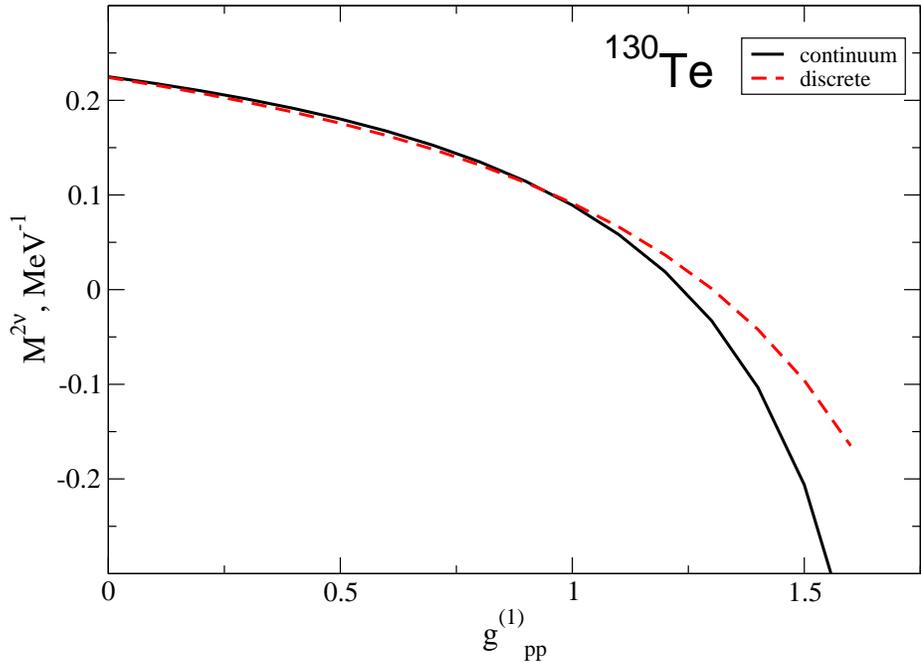}
\end{minipage}
\begin{minipage}[t]{16.5 cm}
\caption{The calculated $g^1_{pp}$-dependence of $M^{2\nu}$ in both discretized and continuum QRPA.}
\end{minipage}
\end{center}
\end{figure}

The calculated values of $M^{0\nu}$ are given in Table I for both versions of the QRPA and $g_A=1.25$.
Note that the two-nucleon short-range correlations are included in the calculations in terms of
the Jastrow function. At the same time, the higher order terms of the nucleon current are not 
considered (they usually reduce $M^{0\nu}$ by about 30\%, see, e.g.~\cite{Rod05}). 

The contributions of the multipoles up to $J^\pi=6^-$ ($L=0\div 5$) are included in the calculations of $M^{0\nu}$.
Note that the QRPA itself as a long-wave approximation is barely suitable to describe 
the multipole contributions with $L>5$ (they contribute in total about 10\% to $M^{0\nu}$), 
they are completely dominated by the short-range behavior of the wave function. 
The partial multipole contributions to the calculated $M^{0\nu}$'s are given in Fig.~2.

\begin{figure}[tb]
\begin{center}
\begin{minipage}[t]{8 cm}
\epsfig{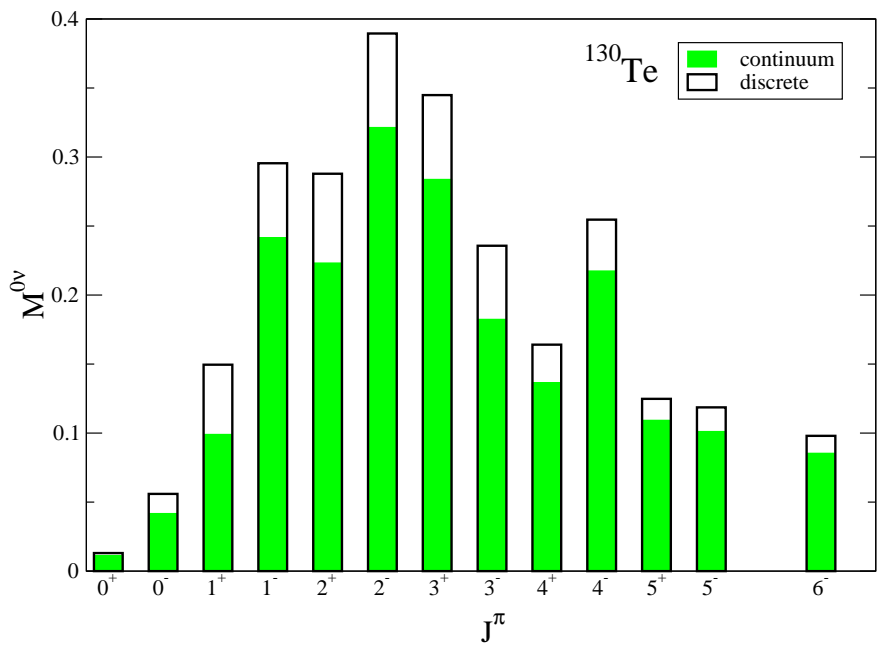}
\end{minipage}
\begin{minipage}[t]{16.5 cm}
\caption{Multipole decomposition of $M^{0\nu}$ calculated within the both versions of the QRPA.
$J^\pi$ is the angular momentum and parity of the intermediate states.}
\end{minipage}
\end{center}
\end{figure}

\section{Conclusions}
In the article a continuum-QRPA approach to calculation of \bb- and \bbb-amplitudes has been formulated.
For $^{130}$Te a regular suppression (about 20\%) of the $(L\ge 2)$-multipole contributons to $M^{0\nu}$ has been found
which can be associated with additional ground state correlations appearing from the transitions to 
collective states in the continuum.
At the same time the total $M^{0\nu}$ for $^{130}$Te gets suppressed by about 20\% as compared to the result of 
the discretized QRPA.
As the nearest perspective we are going to perform a systematic analysis of other double-decaying nuclei within the cQRPA.

\subsection*{\it Acknowledgments}
The work is supported in part by the Deutsche 
Forschungsgemeinschaft (grant FA67/28-2) and 
by the EU ILIAS project (contract RII3-CT-2004-506222).

\end{document}